\documentclass[usenatbib,useAMS]{mnras}

\usepackage{newtxtext,newtxmath}
\usepackage{amsmath}
\usepackage{bm}

\usepackage[T1]{fontenc}
\usepackage{ae,aecompl}

\usepackage{multicol}
\usepackage{graphicx}
\usepackage{pdflscape}
\usepackage{color}
\usepackage{xcolor}

\title[Building Linearly-Evolving Void Catalogues]{How to Build a Catalogue of Linearly-Evolving Cosmic Voids}
\author[S. Stopyra, H. V. Peiris, and A. Pontzen]{Stephen Stopyra$^{1}$%
	\thanks{Contact e-mail: \href{mailto:s.stopyra@ucl.ac.uk}{s.stopyra@ucl.ac.uk}}, Hiranya V. Peiris$^{1,2}$, Andrew Pontzen$^{1}$
	\\
	$^{1}$Department of Physics and Astronomy, University College London, London WC1E 6BT, UK\\
	$^{2}$The Oskar Klein Centre for Cosmoparticle Physics, Department of Physics,\\ Stockholm University, AlbaNova, Stockholm SE-106 91, Sweden}

\date{Accepted XXX. Received YYY; in original form ZZZ}
\pubyear{2020}

\begin{document}
\label{firstpage}
\pagerange{\pageref{firstpage}--\pageref{lastpage}}
 \maketitle

	\begin{abstract}
		\selectfont Cosmic voids provide a powerful probe of the origin and evolution of structures in the Universe because their dynamics can remain near-linear to the present day. As a result they have the potential to connect large scale structure at late times to early-Universe physics. Existing ``watershed''-based algorithms, however, define voids in terms of their morphological properties at low redshift. The degree to which the resulting regions exhibit linear dynamics is consequently uncertain, and there is no direct connection to their evolution from the initial density field. A recent void definition addresses these issues by considering ``anti-halos''. This approach consists of inverting the initial conditions of an $N$-body simulation to swap overdensities and underdensities. After evolving the pair of initial conditions, anti-halos are defined by the particles within the inverted simulation that are inside halos in the original (uninverted) simulation. In this work, we quantify the degree of non-linearity of both anti-halos and watershed voids using the Zel'dovich approximation. We find that non-linearities are introduced by voids with radii less than $5\,\mathrm{Mpc}\,h^{-1}$, and that both anti-halos and watershed voids can be made into highly linear sets by removing these voids.
	\end{abstract}


	\section{Introduction}
	\label{sec:intro}

A large fraction of the volume in the evolved Universe consists of void regions --- areas of significantly lower density than the filaments and clusters where most galaxies are found. When considering the mapping of initial conditions in the early Universe to such regions, the underdensities that evolve into void regions can typically be usefully approximated by linear dynamics for a significantly longer period than overdensities of similar magnitude. If such regions exhibiting linear dynamics can be reliably identified in the evolved Universe, they  provide new routes to extracting reliable cosmological information, suppressing some of the uncertainties and computational expense associated with non-linear evolution. Further, the study of such regions has the potential to provide a sharper view of the pristine early Universe, before information is erased by non-linear evolution.

A substantial literature explores the utility of underdense regions exhibiting linear dynamics as cosmological probes. For example, \citet{10.1111/j.1365-2966.2010.16197.x} constructed an analytic model of void ellipticity using the Zel'dovich approximation \citep{zel1970gravitational}. This can to be used to probe tidal effects \citep{park2007void,lee2009constraining} and to measure cosmological parameters using the Alcock-Paczynski test \citep{sutter2012first,Sutter:2014oca}. \citet{hamaus2014universal} constructed a universal density profile for voids and use the Zel'dovich approximation to predict the expected velocities given this density profile.  This can be used, for instance, to model the redshift space distortions expected around voids \citep{hamaus2015probing}. Furthermore, a good understanding of how void profiles change under different models of cosmology allows for precision constraints on e.g. modified gravity \citep{zivick2015using,falck2018using}. However, such tools are most useful in practice if they are applied to void-like regions with dynamics captured by the Zel'dovich approximation (which is linear in the velocity field and displacements, and mildly non-linear in the density field), since applying them to highly non-linear regions would yield unquantified modelling systematics. 
	
However, many existing methods for identifying voids (in simulations and in observations) are based not on the dynamical properties of cosmic large scale structure, as is the case with the \citet{10.1111/j.1365-2966.2010.16197.x} method\footnote{\citet{10.1111/j.1365-2966.2010.16197.x} used Lagrangian methods to trace galaxies back to their origin in the initial conditions, and then identified voids by searching for maxima of the divergence of the displacement (or velocity) field.}, but rather on the morphological properties of the cosmic web. The majority of the literature relies on ``watershed'' void-finders such as {\tt ZOBOV} \citep{Neyrinck:2007gy}, which is also used as the primary void-finding component of the widely-used {\tt VIDE} package \citep{Sutter:2014haa}. The fundamental idea behind watershed voids is that one first computes a density field from a catalogue of density tracers (which may be halos, galaxies, or simulation particles), and then locates minima of this density field to form the core of low density regions. Adjacent regions are then joined together to form voids by filling outwards with an imaginary rising water level, that incorporates neighbouring regions that the `water' flows into as sub-voids if they are shallower in density, and stops if it encounters a deeper minimum. This approach has been used successfully to extract the void ellipticity distribution and stacked radial density profiles from the \emph{Sloan Digital Sky Survey} \citep{sutter2014voids}. Other authors have used {\tt VIDE} to extract information about cosmological parameters from observations, for example by comparing the void size distribution \citep{Nadathur:2016nqr}, or by performing the Alcock-Paczynski test \citep{Sutter:2014oca}. However, since this morphological definition of voids does not link directly to \emph{dynamical} aspects of void evolution, in general such algorithms will produce void catalogues that are a mixture of regions that are well-described by the Zel'dovich approximation, and those strongly affected by fully non-linear evolution. 
	
\citet{Pontzen:2015eoh} introduced an alternative approach which leads to a simple  dynamical definition of void regions. This relies on comparing a pair of $N$-body simulations which are related by inverting the sign of the Gaussian initial conditions of one simulation with respect to the other, thereby transforming initial underdensities into overdensities and vice versa. By evolving both simulations forward, one can identify voids with \emph{anti-halos} -- regions defined by particles within the original uninverted simulation which end up in halos in the inverted simulation. Since the relationship between halos and initial conditions is well understood, the same analytic methods, such as excursion set models \citep{1974ApJ...187..425P,1991ApJ...379..440B,sheth1999large,Sheth:2003py}, can be used to predict the anti-halo abundance. This is in contrast to watershed voids, where it is difficult to link their abundance to excursion set predictions \citep{Nadathur:2015qua}. While the anti-halo definition is currently difficult to link directly to galaxy catalogues, this situation will soon change through the advent of methods to probabilistically reconstruct the dynamical evolution of large scale structure underlying galaxy surveys, such as {\tt BORG} \citep{jasche2013bayesian}.

In this paper, we will use the Zel'dovich approximation to quantify and compare the dynamical linearity of void catalogues obtained by applying the watershed and anti-halo void definitions to the same $N$-body simulation. Our technique for quantifying {the} dynamical linearity relies on comparing the density field of the fully non-linear simulation to that obtained by extrapolating the initial density field to redshift $z=0$ using the Zel'dovich approximation. By applying this technique to stacked void profiles from the two void catalogues, we will find that the anti-halo definition yields a pure sample of underdense regions accurately described by linear dynamics well into late times, and confirm that the watershed definition can also provide such a sample provided that an appropriate cut is made on the void radius. 
	
In Sec.~\ref{sec:methods} we describe how we construct stacked void profiles, present the simulations used in this work, and  introduce our method for quantifying linearity. In Sec.~\ref{sec:results} we apply this method to void catalogues constructed using the anti-halo and watershed definitions on the same simulation, compare the results, and show how to make radius cuts to select regions with linear dynamics from void catalogues. We discuss the results and future directions in Sec.~\ref{sec:conc}.
	
	\section{Methods}
	\label{sec:methods}
	
	\begin{figure*}
		\centering
		\includegraphics[width=\textwidth]{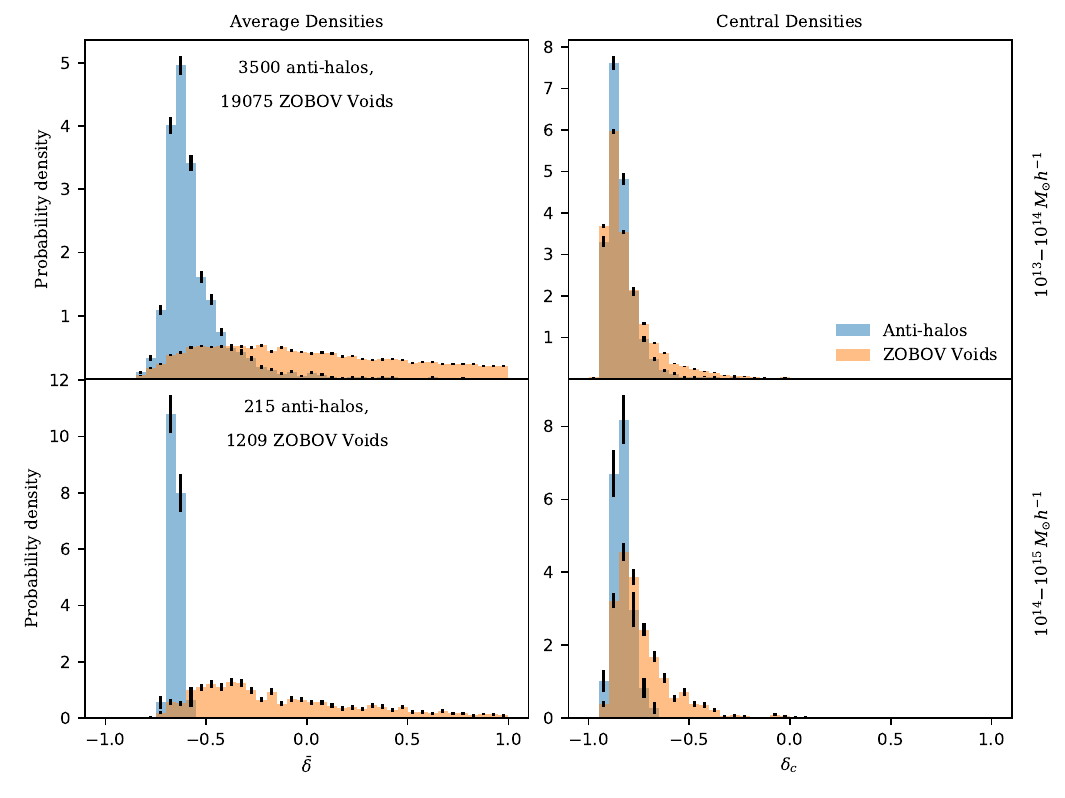}
		\caption{\label{fig:hist1d}The distribution of void densities in the low-mass bin
			($10^{13} \le M/(M_{\mathrm{\odot}}\,h^{-1}) \le10^{14}$, \emph{top row}) and the high-mass bin ($10^{14} \le M/(M_{\mathrm{\odot}}\,h^{-1}) \le 10^{15}$, \emph{bottom row}) for {\tt ZOBOV} voids and anti-halos with (\emph{left column}) average densities $\bar{\delta}$ and (\emph{right column}) central densities, $\delta_c$. {\tt ZOBOV} voids predict large numbers of overcompensated voids with $\bar{\delta} > 0$ that are in the process of being crushed by larger scale overdensities, but have not yet completed this process. In these mass bins, very few anti-halos are above the average density.}
	\end{figure*}

In this work we make heavy use of the radial density profile of a stack of voids: we discuss how these stacks are constructed in Sec.~\ref{sec:definitions}. In Sec.~\ref{sec:simulations} we describe the simulations and explain how we obtain void catalogues from them; the selection criteria applied to the voids identified in the simulations are then presented in Sec.~\ref{sec:subsets}. We outline our method for quantifying linearity of the voids in the resulting catalogues in Sec.~\ref{sec:linearity}.
	
	\subsection{Void Stacking}
	\label{sec:definitions}
	
Stacked void density profiles are key observables used to characterise voids and extract cosmological information. The idea is to take a sample of voids, rescale them by their effective radius, and average the density over the whole set. Such stacks have successfully been used, for example, to perform the Alcock-Paczynski test using voids \citep{sutter2012first}.

	We begin by reviewing how we construct a stacked void profile. The effective radius, $R_{\mathrm{eff}}$, of a void is defined as the radius of a sphere that would have the same volume, $V$, as the void,
	\begin{equation}
	R_{\mathrm{eff}} \equiv \left(\frac{3V}{4\pi}\right)^{1/3}. \label{eq:Reff}
	\end{equation}	
Obtaining $R_{\mathrm{eff}}$ therefore requires us to measure the void volume. The void finders we consider are all able to identify the set of particles in a simulation snapshot that correspond to a particular void. By performing a Voronoi tessellation (which is already accomplished as part of the {\tt ZOBOV} void finder), we can assign a unique volume to each particle that is related to the local density in the vicinity of that particle. This defines the volume, $V_i$, of the $i$th particle. Voronoi cells tessellate by definition, so the total volume of a void can be defined as the sum of the volumes associated with each particle. Since each particle has a well-defined mass, this also simultaneously defines a mass for each void.
	
We can then define the volume weighted density, 
	\begin{equation}
	\rho_{v} = \frac{\sum_i\rho_iV_i}{\sum_i V_i},\label{eq:rhov}
	\end{equation}
which is the mean density of the particles in a void, weighted by the volume assigned to each particle. The advantage of this weighting is that it avoids the estimate being biased by the high density of halos, which contain a large fraction of the particles and mass in a simulation, but only a very small fraction of the volume. An unweighted mean density will be dominated by the particles that lie within halos, and hence fails to measure the density of the void.
	
Following \citet{Nadathur:2014qja}, we then use a volume-weighted void stacking procedure. We average the density in the shell at a given effective radius for each void with a weight given by the shell volume, with a small correction to account for bias\footnote{In practice this does not make a significant difference for our study, since we use simulations with a much higher tracer count than galaxy catalogues.} which otherwise occurs when tracer counts are low. This means the density of a mass shell at a given effective radius is given by
	\begin{equation}
	\bar{\rho}_{\mathrm{{Shell}}} = m\frac{\left(\sum_{i=1}^{N}N_{i,\mathrm{Shell}}\right) + 1}{\sum_{i=1}^{N}V_{i,\mathrm{Shell}}} \, ,\label{eq:stacking}
	\end{equation}
where $N$ is the number of voids in the stack, $N_{i,\mathrm{Shell}}$ the number of particles in the shell for void $i$, and $V_{i,\mathrm{Shell}}$ the (unscaled) volume of shell $i$ for that void. In our case the tracer is dark matter particles which therefore have a constant mass, $m$.
	
We use a different procedure for computing uncertainties on the stacked void profile compared to \citet{Nadathur:2014qja}. For this particular stacking method, they suggested the Poisson error derived from the tracer counts in radial shells, which is appropriate for use in settings utilising sparse tracers of the density field (such as galaxies). In our setting, where $N$-body particles are used to define the voids, and therefore have a much high tracer density, Poisson errors are subdominant in our case to the inherent variability of the void profiles. We discuss this point further in Sec.~\ref{sec:profiles}.

	\subsection{Simulations and Data Processing}
	\label{sec:simulations}
	
We used a pair of simulations previously described in \citet{Pontzen:2015eoh}, where full details can be found. In brief, a pair of $512^3$ simulations denoted $A$ and $B$ were performed, each one with a side $200\,\mathrm{Mpc}\,h^{-1}$ in comoving units. The $A$ simulation was used as the reference universe, in which we would like to identify voids. The $B$ simulation was obtained by inverting the initial conditions of the $A$ simulation. After evolving forward to $z=0$, halos in the $B$ simulation were identified using the {\tt AHF} \citep{AHF} halo finder. The particles associated with each halo in the $B$ simulation were identified in the $A$ simulation using {\tt pynbody} \citep{pynbody}. This defined the $A$ anti-halo associated with each $B$ halo, yielding an anti-halo catalogue for the $A$ simulation. To compare this approach with existing void finders, we ran the {\tt VIDE} code \citep{Sutter:2014haa} which uses {\tt ZOBOV} \citep{Neyrinck:2007gy} as its primary void finder, on the $A$ simulation. In the following, we will refer to the regions in the resulting catalogue as {\tt ZOBOV} voids.
	
For each {\tt ZOBOV} void and anti-halo, the volume was computed using the sum of the Voronoi cells as determined from the output files of {\tt VIDE}. These were then used to compute the effective radii, as well as the volume-weighted barycentres defined by
	\begin{equation}
	\mathbf{x}_{\mathrm{VWB}} = \frac{\sum_i\mathbf{x}_iV_i}{\sum_iV_i},\label{eq:VWB}
	\end{equation}
where $\mathbf{x}_i$ is the position of each particle in the void, and $V_i$  its volume weight as determined from the particle's Voronoi cell. This step was performed accounting for the periodic boundary conditions, ensuring that the resulting centres actually lie within the void for regions that cross the periodic boundary. The volume weighting of the barycentre ensures that it is less susceptible to large fluctuations due to shot noise induced by the location of individual large halos.

After computing the barycentre and effective radius for each void in the catalogues, and assigning each void particle a density using the inverse of its Voronoi volume, $\rho_i = m/V_i$, we then computed the average density and the central density of each void. The average density, $\bar{\delta}$, is defined as the volume-weighted density contrast for all particles making up a void,
	\begin{equation}
	\bar{\delta} = \frac{\rho_v-\bar{\rho}}{\bar{\rho}},\label{eq:delta-bar}
	\end{equation}
where $\bar{\rho}$ is the cosmological average density and $\rho_v$ is given by Eq.~(\ref{eq:rhov}). The central density $\delta_c$ is obtained by computing the volume-weighted density contrast of the $64$ particles\footnote{The precise number of particles is arbitrary, but needs to be sufficient to suppress shot noise in the central density estimate.} closest to the volume-weighted barycentre of the void given by Eq.~(\ref{eq:VWB}). Note that some authors \citep{lavaux2012precision} define this to be the density in a sphere of some fraction, usually 1/2 or 1/4, of the effective radius around the barycentre -- this can lead to particularly sparse and pancake-shaped voids having no particles within such a sphere. Our estimate instead adapts naturally to the sparse distribution of particles at the void centre. For the smallest voids, this may include a significant fraction of the void particles. However, we do not use such small voids in our analysis.

A stacked volume-weighted density profile was then computed for each void by counting the number of particles found in spherical shell bins, and dividing by the sum of their Voronoi volumes as in Eq.~(\ref{eq:stacking}). We used 30 equally-spaced radial bins between 0 and 3 effective radii, so that the position and size of the bins scale with the void size. The resulting tracer counts and shell volumes of all voids were stored for use in computing stacks of arbitrary subsets.
	
\begin{figure}
	\centering
	\includegraphics[width=0.45\textwidth]{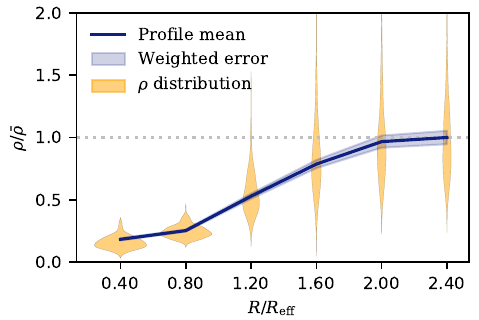}
	\caption{\label{fig:violins} Distribution of the volume-weighted density, $\rho_v$ in radial bins around the centre of the high-mass bin anti-halos. The distribution is non-Gaussian, but the mean (dark, solid blue line) is Gaussian-distributed; the standard deviation of the mean is indicated by the light shaded area.}
\end{figure}

	\subsection{Void Selection}
	\label{sec:subsets}

In order to split the data by void size, we could bin the voids by effective radius. Equivalently, and for easier comparison with \citet{Pontzen:2015eoh}, we instead choose to bin by void mass. We consider a \emph{low-mass} bin in the range $10^{13} \le M/(M_{\mathrm{\odot}}\,h^{-1}) \le 10^{14}$  and a \emph{high-mass} bin defined by $10^{14} \le M/(M_{\mathrm{\odot}}\,h^{-1}) \le 10^{15}$. \citet{Pontzen:2015eoh} showed that smaller masses $M < 10^{13} M_{\mathrm{\odot}}\,h^{-1}$ correspond to regions  that are frequently crushed by larger overdensities, and hence in this range underdensities do not uniquely map to voids. Therefore we do not consider them here. The low-mass bin corresponds approximately to voids with radii $4$--$10\mathrm{\,Mpc}\,h^{-1}$, while the high-mass bin contains voids in the range $10$--$21\mathrm{\,Mpc}\,h^{-1}$. In this $200\,\mathrm{Mpc}\,h^{-1}$ simulation box, anti-halo masses larger than $10^{15}M_{\odot}\,h^{-1}$ are excluded because they are too few in number to obtain good statistics. From the \cite{Tinker:2008ff} mass function, on average $1.4$ such anti-halos or halos are expected in our simulation box (this is identical to the expected number of halos since these voids are too large to have been crushed). In our specific realisation we actually observe 0 anti-halos and 3 halos above this threshold, which is consistent with the mass function at the 95\% level assuming Poisson uncertainties. By contrast, ZOBOV voids are not constructed in a way that matches the halo mass function, and indeed much larger void regions can be found in the ZOBOV catalogues.
	
	\subsection{Quantifying Linearity}
	\label{sec:linearity}
	
	\begin{figure*}
		\centering
		\includegraphics[width=\textwidth]{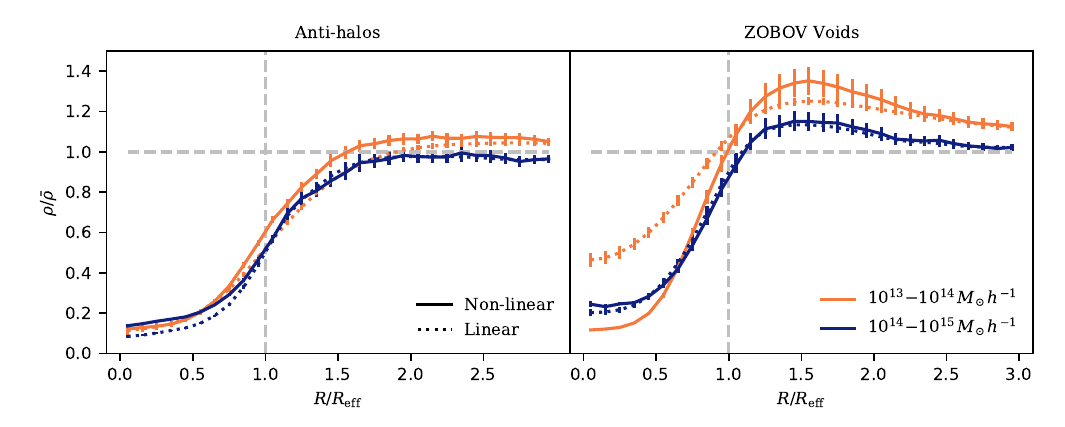}
		\caption{\label{fig:linearVsNonlinearAH} Comparison of stacked density profiles for anti-halos (\emph{left}) and {\tt ZOBOV} voids (\emph{right}) in the low- and high-mass bins. Anti-halos show a broadly self-similar profile that lacks a significant ridge region. The linear profiles (dashed lines) closely match the non-linear profiles (solid lines) for anti-halos. For {\tt ZOBOV} voids on the other hand, this is only true for the high-mass bin; the low-mass linear {\tt ZOBOV} profile differs significantly from its non-linear counterpart. For the {\tt ZOBOV} case, profiles in both mass bins show significant overdense ridges outside the effective radius of the void.}
	\end{figure*}
	
We now wish to quantify the notion of linearity in voids. First, it is necessary to precisely articulate what is meant by ``linearity''. Voids cannot be accurately described by the (Eulerian) linear growth of the density field, since the density contrast extrapolated in this way would become unphysically negative. However, it is possible to obtain a non-negative solution at all epochs that is linear in the displacement, i.e., well-described by first order Lagrangian perturbation theory. From a dynamical perspective, what causes the breakdown of analytic predictability is the onset of \emph{shell-crossing}, which causes first order Lagrangian perturbation theory, also known as the Zel'dovich approximation \citep{zel1970gravitational}, to break down \citep{mo2010galaxy}.  Prior to shell-crossing, the evolution of spherical mass shells can be understood using the spherical collapse model (for halos) and the spherical void model \citep{Sheth:2003py}. 
	
The breakdown of spherical evolution models of collapse and void growth due to shell-crossing is therefore closely related to the breakdown of the Zel'dovich approximation when the first pancakes form. This suggests that by comparing density profiles of voids to the profiles that are predicted by the Zel'dovich approximation, we can quantify whether a given set of voids is well-described by analytic methods and hence can be characterised as \emph{dynamically linear}. Stacks of voids that are strongly affected by non-linearities will have density profiles that differ significantly from their linear profiles. By computing the profiles for different radial bins, we can also determine where non-linear effects are most important in a given stack.
	
In order to implement this procedure, we took the same initial conditions used to compute the $A$ simulation, and evolved them to $z=0$ using the Zel'dovich approximation, which displaces particles according to
	\begin{equation}
	\mathbf{x}(z) = \mathbf{x}_0 + \frac{D(z)}{D(z_0)}\mathbf{\Psi}(z_0,\mathbf{x}_0) \, ,
	\end{equation}
where $D(z)$ is the linear growth factor, $z_0$ the redshift of the initial conditions ($z_0 = 99$ in this case) and $\mathbf{\Psi}(z_0,\mathbf{x}_0)$ the first order displacement field from the initial grid of positions given by $\mathbf{x}_0$. This resulted in a set of linear particle positions, from which we computed linear Voronoi weights. These were used to compute a density field for the Zel'dovich-evolved particles. The $A$ and $B$ simulations were still used to define which particles belong to the {\tt ZOBOV} voids and anti-halos, but for each {\tt ZOBOV} void and anti-halo a second volume-weighted barycentre and effective radius were computed using the linear particle positions. The new Zel'dovich-evolved density field was then used to compute corresponding density profiles for the voids, using the same procedure outlined in Sec.~\ref{sec:definitions} for the non-linear profiles, but using the new barycentres and effective radii. The degree of linearity for a given stack of voids can then be quantified by comparing the Zel'dovich-evolved and simulation-evolved density profiles.
	
	\section{Results}
	\label{sec:results}

We now compare and contrast properties of anti-halos and {\tt ZOBOV} voids in the mass bins defined in Sec.~\ref{sec:subsets}. First, we consider their central and average density distributions in Sec.~\ref{sec:histograms}. Then we move on to considering their stacked density profiles, first discussing how we quantify the uncertainties in Sec.~\ref{sec:profiles} and then considering the implications of the actual profiles in Sec.~\ref{subsec:linearity}. We find that anti-halos are always well described by linear dynamics  in the Zel'dovich sense and show that one can introduce a radius cut to select {\tt ZOBOV} voids exhibiting similarly linear dynamics. We find that there are still qualitative differences between the stacked profiles of anti-halos or {\tt ZOBOV} voids even in such linear samples.
		
	\subsection{Density Distribution of Voids}
	\label{sec:histograms}

We binned the {\tt ZOBOV} voids and anti-halos by their central and average densities: the results are shown in Fig.~\ref{fig:hist1d}. We notice that the {\tt ZOBOV} voids typically have a much longer tail to higher average densities than the equivalent anti-halos. Particularly in the high-mass bin, it is clear that anti-halos are tightly concentrated in both average and central density. For both anti-halos and {\tt ZOBOV} voids, the tail to higher average densities is more pronounced in the low-mass bin, consistent with the findings of \citet{Pontzen:2015eoh} that lower-mass anti-halos are more likely to be crushed due to collapsing overdensities. Further, we see from the right-hand panel for the low-mass bin that  even when they have higher \emph{average} densities,  anti-halos continue to have low \emph{central} densities. We verified that, by contrast, {\tt ZOBOV} voids in the high average density tail also have a tail to higher central densities.

	\subsection{Stacked Density Profiles}
	\label{sec:profiles}
	
	\begin{figure*}
		\centering
		\includegraphics[width=\textwidth]{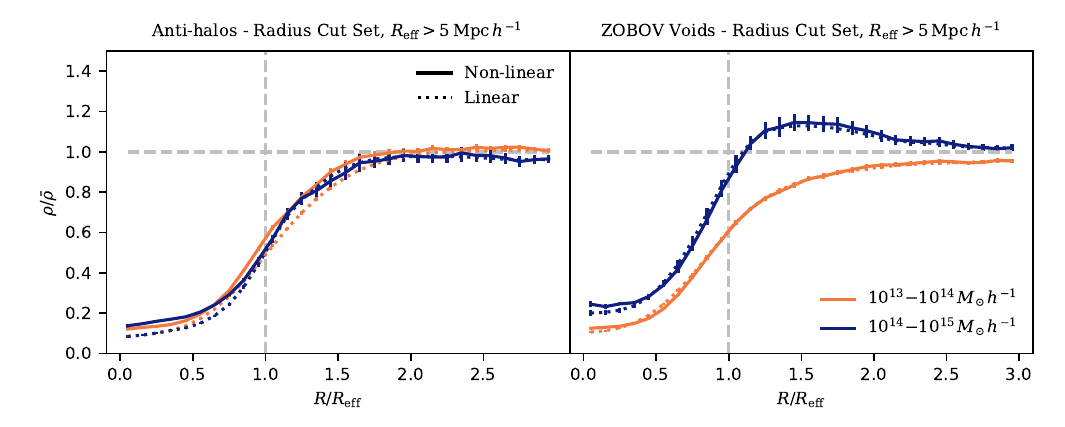}
		\caption{\label{fig:radius_cut}Stacked void profiles in the low and high mass bins with radii $R_{\mathrm{eff}} > 5\,\mathrm{Mpc}\,h^{-1}$. Both the {\tt ZOBOV} voids and anti-halos are close to their linear profiles, indicating that the non-linearities apparent in Fig. \ref{fig:linearVsNonlinearAH} are primarily caused by voids with radii less than $5\,\mathrm{Mpc}\,h^{-1}$. The absence of non-linearities in the { \tt ZOBOV} voids, despite the presence of high-density voids as indicated by Fig. \ref{fig:density_and_radius}, indicates that radius is a more important determiner of non-linearity than average density.}
	\end{figure*}

We now wish to compare stacked density profiles for the two void catalogues in the same mass bins, computed according to the stacking procedure outlined in Sec.~\ref{sec:definitions}. First, we discuss how to characterise the uncertainty in the profiles computed using Eq.~(\ref{eq:stacking}), since (as previously noted) our dominant source of uncertainty is not Poisson errors due to tracer sparsity, but rather the variability of individual void density profiles.
To illustrate this variability, we show the distribution of volume-weighted densities for spherical mass shells within stacked voids in Fig.~\ref{fig:violins}. It is immediately clear that the distribution of densities within radial bins is not Gaussian, as we might expect from the non-linear evolution of the density field. 

However, in order to make a comparison between linear and non-linear evolution, what we wish to characterise is the uncertainty on the \emph{mean} profile in this stack. We note that the volume-weighted mean density for a given radius shell, Eq.~(\ref{eq:stacking}), can be treated as a random variable which is approximately Gaussian-distributed in the limit of a large number of voids in the stack. This follows from the Lyapunov variant of the Central Limit Theorem \citep{BillingsleyPatrick1995Pam}. The error is a generalisation of the $\sigma/\sqrt{N}$ standard error of the mean to weighted means, yielding variance
	\begin{equation}
	\sigma^2_{\mathrm{mean}} = \sum_i(w_i^2)\sigma^2 .\label{eq:variance-on-mean}
	\end{equation}
Here, $w_i = V_i/(\sum_{i}V_i)$ are the weights (normalised to $\sum_{i}w_i = 1$) and $\sigma^2$ is the variance of the profiles arising from the non-Gaussian distribution in Fig.~\ref{fig:violins}. For the case of equal weights $w_i = 1/N$, this reduces to the standard $\sigma^2/N$ variance of the mean. We adopt Eq.~\eqref{eq:variance-on-mean} as the definition of our error bars in all subsequent plots.

	\subsection{Profile Linearity}
	\label{subsec:linearity}
	
	\begin{figure*}
		\centering
		\includegraphics[width=\textwidth]{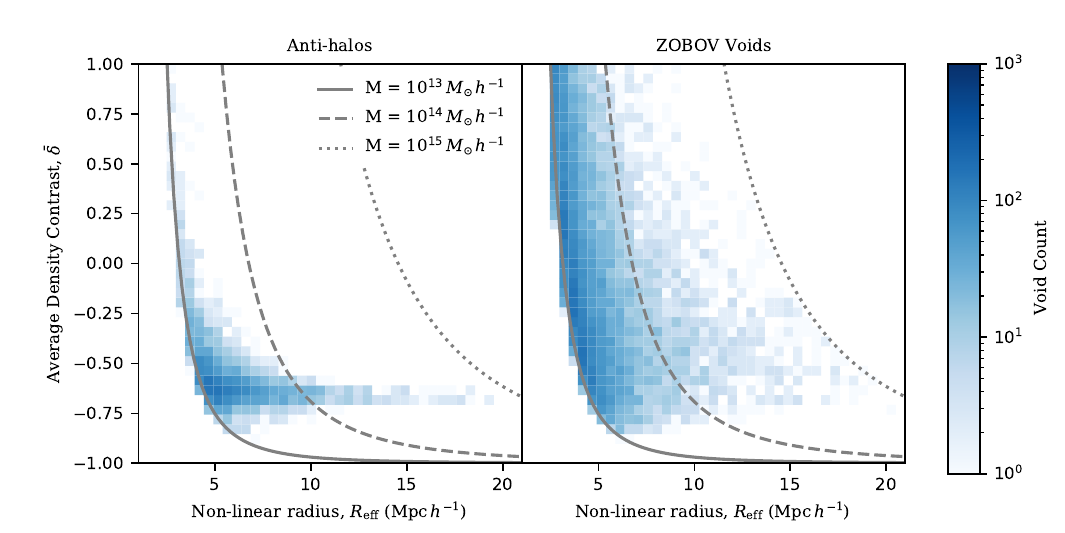}
		\caption{\label{fig:density_and_radius} Distribution of average density, $\bar{\delta}$, and effective radius $R_{\mathrm{eff}}$, for anti-halos and {\tt ZOBOV} voids in the range $10^{13}\mathrm{-}10^{15} M_{\odot}\,h^{-1}$. We over-plot mass contours defined by the relationship between density,  mass and volumes, $\bar{\delta} = M/(V\bar{\rho}) - 1$, where $\bar{\rho}$ is the cosmological average density, $V = 4\pi R^3/3$ the void volume, and $M$ the total void mass. Strictly, this relationship does not have to hold for each individual void, because we use a volume-weighting in our density estimates. However, we verified that it nonetheless accurately divides the histograms up into the mass bins we use for our analysis.}
	\end{figure*}

	Figure~\ref{fig:linearVsNonlinearAH} shows the stacked density profiles for the four void sets under consideration (two mass bins, for both anti-halos and {\tt ZOBOV} voids). We also show the linear prediction in each case as a dotted line, constructed using the same void particles traced to the Zel'dovich-extrapolated snapshot.
	
	Three key differences can be seen between stacked {\tt ZOBOV} voids and anti-halos. First, the former have ``ridge'' regions in their outskirts, which are absent for the latter. Second, anti-halo profiles appear to be well predicted by the linear extrapolation in both mass bins, whereas {\tt ZOBOV} voids are well predicted only in the largest mass bin. Finally, the anti-halo profiles are near-universal with little variation between the two mass bins, while the {\tt ZOBOV} profiles are qualitatively different between bins, for example with a more pronounced overdense ridge in the lower mass bin.  We will now consider the origin of each of these differences, and their inter-relationships.
	
	We verified that the existence of ridges in the {\tt ZOBOV} voids reflects the longer tail to high average densities (Fig.~\ref{fig:hist1d}). This can be confirmed by calculating a {\tt ZOBOV} stack with a cut on the upper bound of the average density $\bar{\delta}$.  In this case, the {\tt ZOBOV} profiles as a function of radius smoothly and monotonically increase to the mean density from below, much like the anti-halo profiles. However, while adjusting the value of the $\bar{\delta}$ cut can qualitatively bring the {\tt ZOBOV} shape into agreement with anti-halo profiles, there is no value which gives quantitative agreement. 
	
	We next consider the correspondence between the non-linear profiles (with density determined from the simulation) and the linear profiles (with density determined from the Zel'dovich extrapolated snapshot). The anti-halos closely match the linear profile at all radii and in both mass bins; the correspondence is particularly close in the outskirts for the high-mass bin, and in the central core for the low-mass bin. By contrast, the {\tt ZOBOV} voids match the linear profile poorly in the low-mass bin, but extremely well in the high-mass bin.
		
	A possible origin for these differences is the process of ``crushing'', which is inherently non-linear since it involves a turnaround in the trajectory of the particles. It was shown by \citet{Pontzen:2015eoh} that smaller voids are more likely to be crushed by $z = 0$, prompting us to check whether the departures from linearity seen in Fig. \ref{fig:linearVsNonlinearAH} are dominated by small radius voids. In Fig. \ref{fig:radius_cut} we show the effect on our stacked profiles of including only those voids with effective radius $R_{\mathrm{eff}} > 5\,\mathrm{Mpc}\,h^{-1}$. The differences between linear and non-linear behaviour seen in the right panel of Fig.~\ref{fig:linearVsNonlinearAH} have now largely disappeared. However, there is still a difference between the shapes of the {\tt ZOBOV} and anti-halo profiles, with the anti-halos remaining universal and the {\tt ZOBOV} voids showing a mass dependence in the prominence of their ridge. (This mass dependence is reversed relative to the uncut case: now the higher mass bin shows a more prominent ridge than the lower mass bin, because the former does not contain any voids small enough to be affected by the radius cut.)  The overall result demonstrates that the origin of the ridge is kinematical rather than dynamical, i.e. it does not rely on the non-linear dynamics of turnaround.
	
	We have so far shown that ridges are linked to the mean density, whereas non-linear dynamics are primarily linked to void radius.   To understand the relationship between cutting in these two variables, we plot the two-dimensional distribution of mean density and radii for the anti-halos and ZOBOV voids respectively in Fig.~\ref{fig:density_and_radius}. Density, radius and mass are interrelated variables, so we overplot lines of constant mass at $10^{13}$, $10^{14}$ and $10^{15}\,M_{\odot}\,h^{-1}$, by combining Eq.~\eqref{eq:Reff} with an expression for overdensity, $\bar\delta = M / (V \bar \rho) -1$. This is approximate rather than exact because the defining relation for $\bar\delta$, Eq.~\eqref{eq:delta-bar}, refers to the volume-weighted mean density $\rho_v$ whereas $M/V$ gives a mass-weighted mean density. Nonetheless we verified that the approximate lines correctly identify the dividing line between our different void bins in the two-dimensional plane  of Fig.~\ref{fig:density_and_radius}.  Comparing the two panels shows that, for antihalos, the average density contrast is almost independent of radius provided one restricts attention to those with $R_{\mathrm{eff}}>5\,\mathrm{Mpc}\,h^{-1}$. The same is not true for ZOBOV voids, which have a tail to high mean densities even at large radii.

	Given the different distributions in key characteristics of {\tt ZOBOV} voids and anti-halos as seen in Fig.~\ref{fig:hist1d} and Fig.~\ref{fig:density_and_radius}, it is likely that the two approaches to void-finding uncover different populations. This is further driven home by the fact that the anti-halo profiles have little mass dependence, while the {\tt ZOBOV} voids are strongly mass-dependent. Fig.~\ref{fig:density_and_radius} gives a possible reason for this inequivalence: all anti-halos above the critical radius of $5\,\mathrm{Mpc}\,h^{-1}$ have a similar average density, since they correspond to halos in the reverse simulation \citep{Pontzen:2015eoh}. By contrast, {\tt ZOBOV} voids are much more diverse. To select a less diverse {\tt ZOBOV} population, and examine whether the correspondence with anti-halos improves,  we can perform a cut retaining only radius $R_{\mathrm{eff}}>5\,\mathrm{Mpc}\,h^{-1}$ voids which are strongly underdense ($\bar\delta < -0.5$). Following this selection cut, we regenerated the stacked profiles, finding that the difference between anti-halos and {\tt ZOBOV} voids persists and moreover, there continues to be a mass dependence in the {\tt ZOBOV} profiles. We conclude that the anti-halos and {\tt ZOBOV} catalogues select different void-like regions.

	\section{Discussion}
	\label{sec:conc}

We have presented an analysis investigating the degree to which cosmic voids identified in simulations exhibit dynamical linearity, where the level of linearity is quantified by the degree to which the Zel'dovich approximation describes the full dynamics of the voids in the evolved simulation. We used this method to compare two different void definitions: anti-halos \citep{Pontzen:2015eoh} and watershed void finders, specifically the {\tt ZOBOV} implementation \citep{Neyrinck:2007gy}. We found that the Zel'dovich approximation can accurately predict the density profiles of anti-halos, whereas an appropriate radius cut is required to select {\tt ZOBOV} voids evolving linearly. We experimented with different cuts, and found that removing voids with radii smaller than $5\,\mathrm{Mpc}\,h^{-1}$ was effective at linearising the stacked void profiles. Even after imposing cuts, the detailed density structure of {\tt ZOBOV} voids and anti-halos is different, with only the former having overdense ridges on their outer edges.

These results can be understood from the physics of void evolution. Spherical models \citep{Sheth:2003py} predict that the time-scale for growth (or collapse) is shorter for small voids than for larger ones. 
The profiles of collapsing voids are inherently hard to predict from linear extrapolations: by studying the 3D evolution of individual such voids over a series of snapshots, we observed that infalling particles often become bound to halos at the void boundary. The capture of infalling particles slows and modifies the collapse process, and cannot be reproduced by the linear physics in the Zel'dovich approximation.  This explains the structure of the non-linear void profiles in Fig. \ref{fig:linearVsNonlinearAH}: the central density is lower than the linear profile, while the ridge density is enhanced.

On the other hand, once a cut in radius is imposed the linear extrapolation becomes accurate even for the {\tt ZOBOV} void stack with a pronounced ridge; that stack includes voids which are in fact overdense. \citet{Nadathur:2016ccl} show that overdense voids must be in the process of collapsing as they constitute minima of the gravitational potential (in contrast to underdense voids which are maxima).  Our results for such large but overdense voids show that, although the collapse is ongoing, it has  not yet had time to reach a non-linear stage.
	
A key attraction of voids is that their quasi-linear evolution may give access to information on relatively small scales (those which, at the power spectrum level, are already non-linear). For example, state-of-the-art galaxy clustering and weak lensing results from the Dark Energy Survey \citep{Krause17DES} do not consider scales below $\simeq 10\,\mathrm{Mpc}\,h^{-1}$, due to the difficulties of modelling non-linear effects below that scale. However, by performing a separate analysis of void regions it should be possible to reinstate some of the lost information. As we have shown, voids with radii $R_{\mathrm{eff}} > 5\,\mathrm{Mpc}\,h^{-1}$ behave very linearly in the sense that they are well-described by the Zel'dovich approximation. Accessing these smaller scales would therefore provide a significant gain, since halving a length scale allows fitting eight times as many modes into the same volume of space, significantly increasing the statistical power. Identifying such voids requires deep data with a sufficiently high density of tracers, but our analysis suggests there is indeed a window where void techniques could give access to small-scale information in the linear regime. Because of the near-linearity, the void profile must retain considerable sensitivity to the correlation function on scales that are even somewhat smaller than the effective radius.  However, detailed study of the information content is beyond the scope of the current work.

Both anti-halos and {\tt ZOBOV} voids can be used to access the information in this regime, provided a suitable cut on radius is included. Anti-halos retain the advantage of a dynamical description which links their abundance and properties to well-studied excursion set methods. However, a drawback is that they are more difficult than watershed voids to identify in observational data. 

This will change in the near future due to the development of powerful methods such as {\tt BORG} \citep{jasche2013bayesian,refId0} which directly fit cosmological simulations to present-day structures traced by galaxies. {\tt BORG} yields three dimensional, probabilistic dynamical reconstructions of the large scale structure underlying galaxy surveys, which can be rewound back to the initial conditions \citep{Leclercq_2015,leclercq2015bayesian,PhysRevD.98.064015,Schmidt_2019}. Such a setting would allow `inverted' simulations to be run starting from initial conditions corresponding to cosmological realisations of our own Universe, allowing the identification of anti-halos within the galaxy survey that was reconstructed. In future work, we will explore this possibility, investigating its potential to produce a pure catalogue of regions which are still undergoing linear evolution in the late-time Universe. 

\section*{Acknowledgements}

The authors gratefully acknowledge useful discussions with Corentin Cadiou, Jens Jasche, Noam Libeskind, Luisa Lucie-Smith, Jenny Sorce and Paul Sutter. This project has received funding from the European Union's Horizon 2020 research and innovation programme under grant agreement No.  818085 GMGalaxies. SS and AP were supported by the Royal Society. HVP was partially supported by the research project grant ``Fundamental Physics from Cosmological Surveys'' funded by the Swedish Research Council (VR) under Dnr 2017-04212. This work used computing equipment funded by the Research Capital Investment Fund (RCIF) provided by UKRI, and partially funded by the UCL Cosmoparticle Initiative.

\section*{Data availability}
The data underlying this article will be shared on reasonable request to the corresponding author.

	\bibliographystyle{mnras}
	\bibliography{voids_paper}
\label{lastpage}
	
\end{document}